\documentclass[twocolumn,aps,pre]{revtex4}
\usepackage{graphicx,color}
\begin{document}
\title{\bf{ Phase-plane analysis of  driven multi-lane exclusion 
models}}     
\author{Vandana Yadav, Rajesh Singh and  Sutapa Mukherji}
\affiliation{Department of Physics, Indian Institute of Technology,
Kanpur-208 016}
\date{\today}
\begin{abstract}
We  show how a  
 fixed point based boundary-layer analysis technique 
can be used to obtain the steady-state particle density profiles of   
 driven exclusion processes on two-lane systems with open boundaries. 
We have considered 
two distinct two-lane systems. In the first,
particles hop on the lanes 
 in one direction obeying exclusion principle  and there is no exchange 
of particles between the lanes. The hopping on one lane is affected 
by the particle occupancies on the other, which thereby introduces  
an indirect interaction  among the lanes. Through a phase plane analysis 
of the boundary layer equation, we  show  why  the bulk density 
undergoes  a sharp change as the interaction between the 
lanes is increased.   The second system 
involves one lane with driven exclusion process and the other with 
biased diffusion of particles. In contrast to the previous model, 
here there is a direct interaction between the lanes 
due to particle exchange between them. 
In this model, we have looked at two possible  scenarios 
with  constant (flat) and non-constant bulk profiles.
 The fixed point based boundary layer method provides  a 
new  perspective on  several aspects including those 
related  to maximal/minimal 
current phases, possibilities of shocks under very restricted 
boundary conditions for the flat profile but over a wide range  of 
 boundary conditions for the non-constant profile.

\end{abstract}
\maketitle
\section{introduction}    
Many unique non-equilibrium 
phenomena such as boundary-induced phase transitions \cite{krug}, 
spontaneous symmetry breaking \cite{evans1}, 
phase separation \cite{jano} are observed 
in  one-dimensional 
driven exclusion processes. These features that are very special 
to systems far-from-equilibrium led to extensive studies 
of variety of driven processes. 
Major  developments in this 
direction  started with the asymmetric simple 
exclusion process (ASEP)  in which particles, after being injected 
at one boundary at a specific rate, hop in a specific 
direction on a finite one-dimensional lattice 
obeying mutual exclusion rule \cite{mukamel,domany}. A steady 
particle current is sustained through  withdrawal of particles 
from the other boundary   at a given rate.  
This simple model was 
 later followed by more general systems  
involving more than one species of particles \cite{evans1,popkov1,
smfixedpt}, more complicated 
lattice \cite{popkov2,peschel,klumpp,harris,evansprl,evans2,melbinger} 
 or dynamics
of particles \cite{hager}  etc. All these models have a few 
common, basic  features such as  
hopping of particles 
with bias in one direction and hence a nonzero particle flux, mutual 
exclusion between the particles and 
open boundaries with particle injection  and withdrawal 
at given  rates.  In contrast to equilibrium systems, 
these one-dimensional, driven, many particle systems exhibit 
boundary induced phase transitions in their steady-state.
  In various phases, 
density profiles have distinct shapes which, for a 
given process, are completely dependent on the boundary rates. 
The boundary rates, therefore, are  the most natural  variables 
for the phase-diagram 
representing the phases and phase transitions.



In order to characterize the phase transitions,  it often appears 
convenient to look at the steady-state density profile which describes  
the average particle occupancy of various sites.  
Density profiles  have interesting shapes 
with  extended bulk parts and 
one or more boundary layer parts which are narrow regions 
over which the density varies rapidly.  
The nature  and location of the bulk, as well as 
 the boundary layer parts of the  profile,
 change as the boundary rates change.  
For example, the boundary layers 
may be located near one or both the boundaries or may appear 
in the interior  of the lattice. The latter one, commonly known 
as a shock, separates high and low density bulk regions 
 \cite{parameg}.  That the boundary layer need not be 
confined to the boundary alone and the  formation of a 
 shock can be well characterized through a 
deconfinement transition of the boundary layer from the 
boundary have been shown in \cite{smbjphys,smbreview,dualjphys}.  
In the next section, we 
shall provide a more technical  description of the boundary layer 
in terms of an appropriate differential equation.

 Recent studies  \cite{smbjphys,smbreview,dualjphys} 
show that the information about the bulk density 
can be obtained by studying the  boundary layer parts 
of the density profile.  Although methods of   boundary layer 
analysis \cite{cole} 
allow us to find the analytical expressions of the 
boundary layers \cite{smbjphys}, 
in many cases, it becomes technically challenging to obtain 
analytical expressions for the boundary layers. However, since the 
boundary layers are expected to merge to the bulk in the appropriate 
limit, one may obtain information about the bulk density
by studying the  fixed points of the differential equations 
describing the boundary layers. For constant bulk profiles, 
this method is especially  powerful  since, in this case, 
 the bulk 
densities must be the same as  the fixed point values 
 of the boundary layer equation.

In the  present work, we apply   fixed point based boundary layer 
 method \cite{smfixedpt} to  two driven many-particle 
systems, 
each composed of  two one-dimensional lattices or lanes.
Every lane has $N$   sites  and particles  can 
hop to the neighboring sites on the lane  obeying specific rules  
that are mentioned in detail in the respective sections. The  
particle dynamics of the 
two models considered here  differ due to the following reasons. 
In the first model, particles do not hop from one lane to the 
other but the hopping rate on one lane changes based on the 
particle occupancy of the neighboring sites on  the other lane 
\cite{popkov2,peschel,melbinger}.
In contrast,  the 
second model involves two lanes  which can mutually exchange 
particles.  While particles on one lane  undergo asymmetric 
simple exclusion process with   hopping in a specific direction, 
those  on the other lane  undergo  biased 
diffusion \cite{klumpp,evansprl,evans2}. Such 
driven systems have 
 similarities with intracellular transport processes
in which molecular motors, receiving energy from the hydrolysis 
of adenosine triphosphate (ATP),  move in a particular direction along 
microtubules. 
The first model may 
mimic the situation where a  molecular motor moving along one channel 
with a large cargo attached to it creates an obstruction  
for the motion of  other molecular motors 
 on the  neighboring channel. 
A second lane with particle diffusion on it  has been introduced 
earlier to represent the environment in which molecular motors
diffuse during the period when they are not attached to the 
microtubule \cite{klumpp}.    
More recently, similar two-lane models 
have been introduced to describe extraction 
of membrane tubes by molecular motors \cite{evansprl,evans2}.
In both the models that  we study here, lanes are coupled 
to boundary reservoirs that maintain specific densities 
at the two ends of a lane.
Our primary aim is to illustrate how the method 
can be used to predict steady-state density profiles under 
different boundary conditions 
 and to quantify various features of  the boundary layers
including the  height of the shock,
approach to the bulk etc.

The motivation behind this study stems from the fact that 
till now there exists  no general framework to study 
phase transitions of this small class of 
non-equilibrium  systems.  Although this is a much bigger issue 
concerning the  entire subject of non-equilibrium statistical mechanics,
lack of a general framework even for these driven systems is 
surprising.  
Due to the presence of more than one density variable, 
driven multi-lane systems 
\cite{popkov2,peschel,klumpp,harris,evansprl,evans2,melbinger}, 
in general, are technically challenging. 
Previous studies on  
 two lane systems, with particle occupancy on one lane 
affecting the hopping on the other, show that this issue requires 
a generalization of the extremal current principle \cite{melbinger} 
which has 
been used earlier to predict  the phase diagram of 
a single lane ASEP involving a single density variable 
\cite{popkovex,hager}.  Inadequacy 
of mean-field analysis for certain cases 
has motivated development of cluster 
approximation \cite{peschel}
 and all these studies have revealed existence of 
new interesting phases including symmetry breaking. 
  Studying the 
stability properties of the bulk plateaux, a variety of 
phase diagrams has also been found for systems with 
 ASEP on one lane coupled 
to different kinds of particle's motion on the other lane \cite{evans2}.
All these observations motivate us to verify the versatility of the 
boundary layer based method to this class of problems. The present 
paper is restricted to  only two specific models from this class. 
It would be interesting  to apply this method   
to study phase diagrams of other variants  of two lane systems.
This  will be discussed in  our future publications.

 The present method appears  to be general since  it 
does not rely on any  explicit analytical 
solution of the density profile, yet it provides  
a lot of physical insight regarding  the location of the 
boundary  layer, value of the bulk density, nature of the phase transitions 
etc; all these
 obtained analytically through a phase-plane analysis of the 
differential equation describing the boundary layer. 
We predict the shape of the entire density profile from the fixed points 
and the phase portraits of the boundary layer equation.
Most importantly, we find that   various 
known  results, obtained   
through development of different methods and hypotheses including 
extremal current principle, follow as natural consequence of the 
phase-portrait of the  boundary layer differential equation.
There are resemblances between the present  method and the extremal current 
principle but it would require more work to establish a direct connection 
which might provide a natural basis for the hypothesis 
made for the  extremal current principle.  

 The stochastic dynamics of  particles undergoing ASEP 
 can be described through 
discrete master equations. For large $N$ ($N\rightarrow \infty$), 
and small lattice 
spacing, $a$ ($a\rightarrow 0$) with $Na$ finite, 
we may go over to a continuum limit 
in which a lattice site $i$  is replaced by a continuous 
 position variable $x=i/N$.  In the continuum, long time, long 
length scale  limit (the so called hydrodynamic limit), 
a statistically averaged master equation appears like a 
continuity equation describing the time-evolution of a density 
variable, say, $\rho(x,t)$,  in terms of the  particle current $j$.  
 Since bulk phase transitions are large length scale  phenomena 
and only certain gross features are crucial at these length scales, 
a continuum formulation is expected to be sufficient for our 
purpose. Our boundary-layer method is applicable to the steady-state 
version of such continuum equation.
 It is important  to note  
that this continuum approach, however,   produces a narrow 
boundary layer whose width varies with $N$. 
This will be shown  explicitly in the next section by obtaining  
the boundary-layer solution  for the simplest ASEP model
introduced at the beginning.

An explicit derivation of the steady-state 
hydrodynamic equation requires obtaining the stationary flux, 
$j(\rho)$ from the microscopic   dynamics 
 of the model. For the first model, 
this derivation is simple due to  spatially uncorrelated 
nature of the steady-state \cite{popkov2} and  
for the second model, we use a mean-field current-density relation.  
Since our primary aim is to elucidate how the fixed point based   
boundary-layer method works, we have used mean-filed approximation 
for convenience. It will be clear from the analysis 
that the method is robust in the sense that 
it works equally well for other hydrodynamic equations
 irrespective of the approximation that has been used 
to derive these.
Hence, the outcome of this method is exact to the extent 
to which the starting hydrodynamic equation is exact.
A phase-plane analysis for 
the  first model shows  that, for certain boundary conditions,
the  density profile is strongly 
 influenced by a saddle fixed point of the boundary
layer differential equation and finally, as a consequence of this,
the bulk profile  changes 
 drastically to a new value  as the interaction between the 
lanes is increased.  
The second model provides an interesting fixed point  diagram 
with saddle-node bifurcations of the fixed points \cite{strogatz}
of the boundary layer equation. These bifurcations appear at special 
densities that correspond to the  maximum or  minimum of the 
particle current and  maximal/minimal current phases 
appear as natural consequences of the flow trajectories 
towards these special densities. 

In section II, we discuss a few general properties of the boundary layer 
solution. Section III presents  the phase plane analysis of the first model. 
In section IV, we consider  the second model with two distinct 
cases of constant and non-constant bulk profiles.
We summarize our main findings in section V.

 \section{Boundary layers}
In order to illustrate some of the basic features of the boundary layer,
here we choose   the simplest model of ASEP in which 
particles hop in a particular direction on one lane
obeying the exclusion rule. In addition, we also include processes that 
involve adsorption and evaporation of particles to and from the lane 
at rates proportional to $\omega_a$ and $\omega_d$, respectively.
The lane  is coupled to boundary reservoirs which maintain fixed 
particle densities, $\rho_l$ and $\rho_r$, at  left and right 
boundaries of the lane.  
For the following equation, equal particle adsorption and evaporation 
rates are assumed.    
In a continuum mean-field description with the lattice size 
scaled to unity ($Na=1$), 
the density variable $\rho(x,t)$ satisfies 
the differential equation 
\begin{eqnarray}
\frac{\partial \rho}{\partial t}= 
\epsilon\frac{\partial^2\rho}{\partial x^2}+(2\rho-1) 
\frac{\partial\rho}{\partial x}+
\Omega(1-2\rho). 
\label{hydro}
\end{eqnarray}
Here $\Omega=\omega_a N=\omega_d N$ and $\epsilon=\frac{1}{2N}$ 
is a small parameter.
In the absence of adsorption and evaporation processes,
the hydrodynamic equation  with only  hopping of particles 
can be expressed  in the form of a continuity equation 
$\frac{\partial \rho}{\partial t}=
-\frac{\partial }{\partial x}(-\epsilon
\frac{\partial \rho}{\partial x}+ 
j_\rho),\label{simplecont}
$
where $j_\rho=\rho(1-\rho)$ is the particle current associated with the 
hopping process. In addition, there is a diffusive  current which appears 
along with a prefactor, $\epsilon$, which becomes small as 
$N\rightarrow \infty$. 
In order to obtain the steady-state  density profile, 
we need to solve 
\begin{eqnarray}
\epsilon\frac{d^2\rho}{dx^2}+(2\rho-1) \frac{d\rho}{dx}+
\Omega(1-2\rho)=0 \label{hydro2}
\end{eqnarray} with
boundary conditions $\rho(x=0)=\rho_l$ and $\rho(x=1)=\rho_r$.   

\begin{figure}[htbp]
  \begin{center}
   \includegraphics[width=2.5 in, clip]{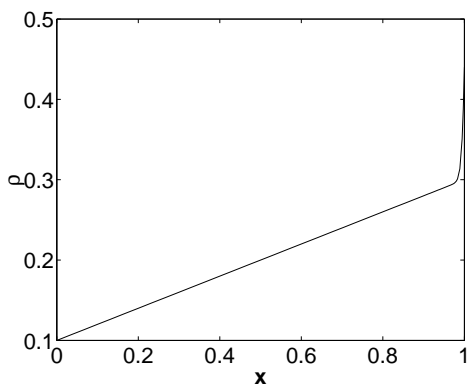}
    \caption{A typical density profile in the simplest ASEP 
model considered in this section.}
\label{fig:examplebl}
  \end{center}
\end{figure}

 In the limiting case, $\epsilon= 0$, equation 
(\ref{hydro2}) becomes a first order equation and its solution 
cannot, in general, 
satisfy two boundary conditions. The vanishing higher-order 
derivative term, also known as the regularization term, 
helps avoiding the singularity in the differential 
equation. In these problems,  
boundary layers are expected to appear near either one of the 
boundaries or in the interior of the lane. 
Based on the values of the boundary densities, different  
solutions arise and the techniques of the boundary-layer analysis 
allow one to obtain uniform approximation to the 
 solution of (\ref{hydro2}) order by order in $\epsilon$ 
under given boundary conditions.
 For very small $\epsilon$, the second order derivative term 
 of equation (\ref{hydro2}) can be neglected. 
The solution of the resulting first order equation  describes 
the major part of the density  profile. This solution, known as 
the outer solution in the boundary-layer language,  is referred 
in the following as the bulk solution.  For this particular 
example, the bulk solution is   
\begin{eqnarray}
\rho_{\rm o}(x)=\Omega x+C_0
\end{eqnarray} with $C_0$ being the integration constant 
whose value can be  determined from the boundary condition that 
this solution satisfies. 
For example, for 
 a  density profile  with the bulk part  satisfying the 
boundary condition at $x=0$, $C_0=\rho_l$.
It is possible that along with this bulk solution, 
a boundary layer  appears near $x=1$ (see figure \ref{fig:examplebl} for 
a typical density profile that appears in the low density phase for the 
ASEP under consideration \cite{parameg,smbjphys}).
The  
 boundary layer here satisfies the boundary condition at $x=1$
and  merges to  the 
bulk solution at  the other end. 
In order to satisfy two conditions,  the 
second derivative term of (\ref{hydro2}) becomes necessary for the 
description of the boundary layer. Thus, in general, higher 
order derivative terms 
dominate the behavior in the boundary layer.    
To focus on the boundary layer, we 
introduce a rescaled variable $\tilde x=\frac{x-x_0}{\epsilon}$, where
 $x_0$, which is arbitrary at this stage, specifies the location of the 
boundary layer after appropriate  boundary conditions are implemented.
In terms of this rescaled variable, (\ref{hydro2}) appears as 
\begin{eqnarray}
\frac{d^2\rho}{d{\tilde{x}}^2}+(2\rho-1) \frac{d\rho}{d\tilde x}+
\epsilon\Omega(1-2\rho)=0. 
\label{hydro2'}
\end{eqnarray}
The $O(\epsilon^0)$ solution of the boundary layer can be obtained 
from 
\begin{eqnarray}
\frac{d^2\rho}{d{\tilde{x}}^2}+(2\rho-1) \frac{d\rho}{d\tilde x}=0.
\label{blayersimple}
\end{eqnarray}
Since, the boundary layer is narrow, this also implies that at this 
order, the particle non-conserving processes have negligible influence 
on the boundary layer. 
Irrespective of their appearance in the interior or the boundary 
of the lane, solution of such equations are, in general, referred 
 as the boundary layer solutions in the following. If 
the boundary layer appears near $x=1$, we expect the boundary layer 
solution  to merge to the bulk at $\tilde x\rightarrow 
-\infty$ and satisfy the 
boundary condition $\rho(\tilde x=0)=\rho_r$. 
For boundary layers appearing  in the interior of a lane 
separating high and low density bulk solutions,
we expect the boundary layer solution 
to merge to   appropriate bulk densities in  
the $\tilde x\rightarrow \pm\infty$ limits.  

Integrating (\ref{blayersimple}) once, we have 
\begin{eqnarray}
\frac{d\rho}{d{\tilde{x}}}+(\rho^2-\rho)=c_0, \label{blayersimple2}
\end{eqnarray}
where $c_0$ is the integration constant. The saturation of the 
boundary layer to the bulk density, $\rho_b=\rho_o(x=1)$, 
requires $c_0=\rho_b^2-\rho_b$.
This relation between $c_0$ and $\rho_b$ fixes the 
 range of physically acceptable values of $c_0$. 
 Approach to different bulk values can be also
understood by obtaining phase-portraits of such equations for 
different values of $c_0$ in this range. 
Fixed points, $\rho^*$, of this equation are the solutions of the 
equation ${\rho^*}^2-\rho^*-c_0=0$.
It often appears convenient to plot these fixed points
as functions  of $c_0$.
Obtaining the stability properties of the fixed points, 
one may indicate on this diagram, how, starting 
from a given initial density, 
the solution flows towards or away from   a given 
fixed point. Such diagrams will be referred in the following as 
 fixed point diagrams.

Instead of elaborating on the fixed
 point diagram of this model in detail, 
we discuss some of the properties of the boundary layer by
  focusing  on the  density profile of figure (\ref{fig:examplebl}). 
Clearly, the boundary layer 
appearing at $x=1$ has to merge to the bulk density, $\rho_b$. 
We expect the bulk-density $\rho_b$ to be an unstable fixed point 
of the boundary layer to which the boundary layer merges as 
$\tilde x \rightarrow -\infty$. Whether the other end of the 
boundary layer ($\tilde x\rightarrow \infty$ limit) approaches 
a fixed point or flows indefinitely is  decided by the nature of  
the boundary-layer equation and can be seen clearly from the 
fixed point diagrams.
In this particular case,  boundary layer equation 
(\ref{blayersimple}) has simple  solutions 
\cite{smbjphys,smbreview}
\begin{eqnarray}
\rho_{bl}(\tilde x)=\frac{1}{2}+\frac{(1-2\rho_b)}{2} 
\tanh(\frac{\tilde x}{2w}+\xi) \ \ {\rm and}\\
\rho_{bl}(\tilde x)=\frac{1}{2}+\frac{(1-2\rho_b)}{2} 
\coth(\frac{\tilde x}{2w}+\xi),
\end{eqnarray}  
where $\xi$ is a constant that gives the center of the boundary 
layer and $w=\frac{1}{1-2\rho_b}$ is the 
width of the boundary layer. While $w$ describes the approach of the 
boundary layer to the bulk asymptote, $\xi$
helps visualizing the shock formation. These solutions also show 
how the boundary layers scale with the system size, $N$. 
The  boundary layer  
of figure (\ref{fig:examplebl}) is described 
by the $\tanh$  solution  which satisfies the boundary condition 
at $x=1$  before approaching its stable fixed point, $1-\rho_b$, 
as $\tilde x\rightarrow \infty$, a limit that 
 goes beyond the physical size 
of the lane. 
In case the boundary layer  has a  stable fixed point, which is 
the case for the  $\tanh$ type boundary 
layer shown in figure (\ref{fig:examplebl}), 
one may expect to see a shock as the 
boundary condition at $x=1$ is appropriately adjusted. 
The boundary layer 
just deconfines from the $x=1$ boundary and enters into 
 the bulk in the form of 
a shock when  the stable fixed point of the boundary layer 
 is exactly the same as the boundary condition at $x=1$.  
At this instance, the $\tilde x\rightarrow \infty$ end 
of the boundary layer is just in the physical region.  
Now, if the boundary condition at $x=1$ 
 is raised to a slightly  higher value, the 
 boundary layer, that has already 
reached its saturation  at $x=1$ cannot anymore  satisfy the 
boundary condition. 
It is, in this case, that the boundary-layer 
deconfines and enters into the bulk as a shock with the 
$\tilde x\rightarrow \infty$ merging to another high-density 
bulk solution. This example of ASEP presents a simpler problem 
since, in this case, 
 one can explicitly solve the  boundary layer equations. 
Fixed point diagrams are especially useful for those cases 
for which such exact solutions cannot be obtained.

\section{Model-1: Two-lane system without particle exchange}
\begin{figure}[htbp]
  \begin{center}
   \includegraphics[width=3.5 in, clip]{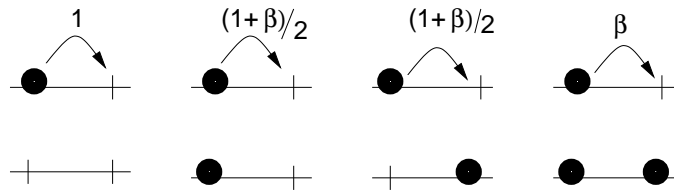}
    \caption{The hopping  rates on one lane
 depend on the total particle occupancy of the  
neighboring sites on the other lane. On a lane, a particle 
hops to the empty, forward site at rates $1$, $(1+\beta)/2$ and 
$\beta$ if the neighboring sites on the other lane are empty, 
half-filled and completely filled respectively. }
\label{fig:model1fig}
  \end{center}
\end{figure}

 In this model, we have two lanes on which particles hop 
unidirectionally at rates as shown in figure (\ref{fig:model1fig}). 

Since the  hydrodynamic   equations  have been already 
derived from the microscopic dynamics of the model, 
we  quote these equations 
from the previous literature \cite{popkov2}. 
In the steady-state, these  equations are 
\begin{eqnarray}
-\frac{\partial j_\rho}{\partial x} +
 \epsilon\frac{d}{dx}[(1-m \sigma)\frac{d\rho}{dx}]=0,\ \  {\rm and} 
\label{cont1}\\
-\frac{\partial j_\sigma}{\partial x}+ 
\epsilon\frac{d}{dx}[(1-m\rho)\frac{d\sigma}{dx}]=0, \label{cont2}
\end{eqnarray}
where $\epsilon=\frac{1}{2N}$ and $\rho$ and $\sigma$ are the 
average  densities on the lanes  with the corresponding currents 
$j_\rho=\rho(1-\rho)[1-(1-\beta) \sigma]$ and  
$j_\sigma=\sigma(1-\sigma) [1-(1-\beta) \rho]$. Here, $\beta$ 
measures the strength of the interaction between the two lanes 
with $\beta=1$ representing the non-interacting case.
 From now onwards, 
we use $m=1-\beta$.  These equations are to be supplemented 
 with  the boundary conditions   
($\rho(x=0)=\rho_l$, $\rho(x=1)=\rho_r$) and ($\sigma(x=0)=\sigma_l$, 
$\sigma(x=1)=\sigma_r$) at the two ends of the lanes.  
Instead of considering  nonlinear 
regularization terms, we follow 
a phenomenological approach and choose simpler 
regularization terms of the form   $\epsilon \frac{d^2 \rho}{dx^2}$  and 
$\epsilon \frac{d^2\sigma}{dx^2}$. After studying the steady-state for 
this simpler situation, we argue that no new feature emerges with 
the actual regularization terms.

Hydrodynamic equations that we study in the following are,
\begin{eqnarray}
 -\frac{\partial j_\rho}{\partial x}+
\epsilon\frac{\partial^2\rho}{\partial x^2}=0 \ {\rm and}\label{naive1}\\ 
-\frac{\partial j_\sigma}{\partial x}+
\epsilon\frac{\partial^2\sigma}{\partial x^2}=0.\label{naive2}
\end{eqnarray}
Equations (\ref{naive1}) and (\ref{naive2})
 admit constant solutions which 
correspond to constant bulk profiles.  
For the boundary-layer solutions, we re-express  these  equations 
in terms of  $\tilde x$  and integrate 
once to obtain   
\begin{eqnarray}
\frac{\partial \rho}{\partial \tilde x}=\rho(1-\rho)(1-m \sigma)+
c_1, \label{bl1}\\
\frac{\partial \sigma}{\partial \tilde x}=\sigma(1-\sigma)(1-m \rho)+d_1.
\label{bl2}
\end{eqnarray}
Here $c_1$ and $d_1$ are the two integration constants.
The saturation of the boundary layers to  the bulk densities $\rho_b$ 
and $\sigma_b$ is ensured through the choice
\begin{eqnarray}
c_1=-\rho_b(1-\rho_b)(1-m \sigma_b),\label{const1}\ {\rm and}\ \\ 
d_1=-\sigma_b(1-\sigma_b)(1-m \rho_b).\label{const2}
\end{eqnarray}

 To begin with, we consider a situation 
where a  density  profile has  
a constant bulk part satisfying 
one boundary condition and a boundary-layer 
part satisfying the other boundary condition.  
As discussed earlier,  one may conclude that the 
bulk density value which is also a boundary 
density at one end, is a fixed point of the 
boundary layer equation. 
The boundary layer, in this case, is the solution of (\ref{bl1}) 
and (\ref{bl2}) that 
starts from  an initial density, which, here,
 is the boundary condition satisfied
by the boundary layer,  and approaches the fixed point.  
Hence, for a set of boundary conditions, we may 
find  out all the fixed points of 
the boundary layer equations with $c_1$ and $d_1$ fixed by assuming 
bulk densities 
to be same as one set of the boundary values. Whether this  choice 
of bulk density and the corresponding boundary layer form 
an acceptable solution for the density profile, depends on the 
stability properties of  fixed points and the phase-plane trajectories.
Although settling this issue 
 is much simple when the boundary layer equation 
has only stable or unstable fixed points, this is not so when 
the boundary layer equation has a saddle fixed point 
in addition to stable or unstable fixed points.  
In order to see these features, the  most general 
approach would be to obtain the fixed point diagram as functions 
$c_1$ and $d_1$. For finding out the fixed points, 
$\rho^*$ and $\sigma^*$,
one has to solve algebraic equations, 
$\rho^*(1-\rho^*) (1-m \sigma^*)+c_1=0$ 
and $\sigma^*(1-\sigma^*)(1-m \rho^*)+d_1=0$. Solving a fifth order 
polynomial equation for either $\rho^*$ or $\sigma^*$, one 
may find five fixed points of which only real, positive  fixed points 
of values $\rho^*({\rm and}\ \sigma^*)\le 1$ are the physically 
acceptable ones. Clearly, this would  require a multi-dimensional 
parameter space on which the topology of the manifolds describing 
the fixed-point would be displayed.
Instead of this detailed fixed point diagram, one may also find 
fixed points of (\ref{bl1}) and (\ref{bl2}) with values 
of $c_1$ and $d_1$ obtained for all possible  combinations  
of $\rho_b$ and $\sigma_b$. This allows us to 
 have a generic picture for the 
number of physically acceptable 
fixed-points and their stability properties on the entire 
$\rho_b-\sigma_b$ plane.

Instead of providing a detailed picture,  
we consider a specific case of 
bulk densities $\rho_b=0.4, \ {\rm and}\ \sigma_b=0.01$. 
It turns out that this  corresponds
 to the same shape of the density profile 
we have just discussed. 
Out of  five sets of fixed points 
of the  boundary layer equation,  
only two  sets are physically meaningful and these fixed points 
 govern the phase-plane 
trajectories of the boundary layer solutions. The other fixed points 
are not crucial for our analysis since these  are either 
imaginary or  are unphysical  (fixed point values larger than $1$). 
Table 1, in the following, provides a
list of  values of $c_1$,  $d_1$,  two physically acceptable sets of 
fixed points $(\rho_1^*,\sigma_1^*)$ and $(\rho_2^*,\sigma_2^*)$,
 and the corresponding  sets of  eigenvalues $(\lambda_{11},\lambda_{12})$
and $(\lambda_{21},\lambda_{22})$ 
  for  different  values of $m$.
\begin{widetext}
\begin{tabular}{|c|c|c|c|c|c|c|}\hline
m & $c_1$ & $d_1$ & $(\rho_1^*,\sigma_1^*)$ & $(\rho_2^*,\sigma_2^*)$ &
$(\lambda_{11},\lambda_{12})$ &$(\lambda_{21},\lambda_{22})$\\
\hline
0.48 &-0.23884 &-0.00799 & (0.4,0.01) & (0.5992,0.01134) &(0.7927,0.1981) 
&((0.698,-0.199)\\
0.51 &-0.23877 &-0.00788 & (0.4,0.01) & (0.5991,0.0114) & (0.781,0.198) &
(0.6808,-0.1997)\\
0.58 & -0.238608 &-0.00760 & (0.4,0.01) & (0.598, 0.01178) & (0.754,0.197) 
& (0.638,-0.195)\\
0.68 &-0.238368 & -0.0072072 & (0.4,0.01) &(0.5980,0.0122) 
& (0.71556, 0.1965) & (0.5794,-0.20) \\
\hline
\end{tabular}\\ 
\end{widetext}

\begin{figure}[htbp]
  \begin{center}
   \includegraphics[width=3.5 in, clip]{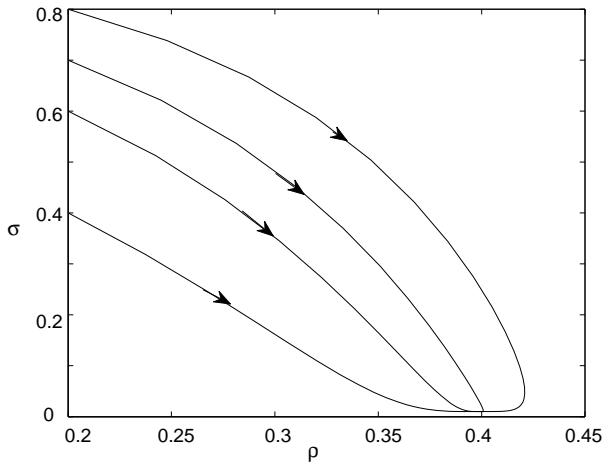}
    \caption{Phase-plane diagram on the $\rho-\sigma$ plane for different 
initial conditions with  $m=0.4$. 
Initial conditions $(\rho(\tilde x=0),\sigma(\tilde x=0))$ 
as we move towards the outermost line 
are $(0.2,0.4)$, $(0.2,0.6)$, $(0.2,0.7)$, $(0.2,0.8)$.
The arrows on the lines, 
indicate the direction of more negative values of $\tilde x$.}
\label{fig:fixedmpt4}
  \end{center}
\end{figure}
\begin{figure}[htbp]
  \begin{center}
   \includegraphics[width=3.0 in, clip]{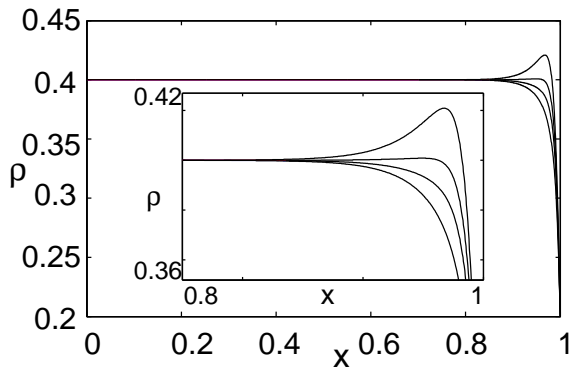}
    \caption{Density profiles ($\rho$)  plotted for $m=0.4$ 
and  left boundary conditions $(\rho_l,\sigma_l)=(0.4,0.01)$. Right boundary 
conditions, here, are 
 same as the initial conditions of figure \ref{fig:fixedmpt4}.
 The lower most profile corresponds to $(\rho_r,\sigma_r)=(0.2,0.4)$.
The rest follow the same order as specified in figure \ref{fig:fixedmpt4}.
All the plots are obtained with $\epsilon=0.006$.}.
\label{fig:fullden1}
  \end{center}
\end{figure}
\begin{figure}[htbp]
  \begin{center}
   \includegraphics[width=3.0 in, clip]{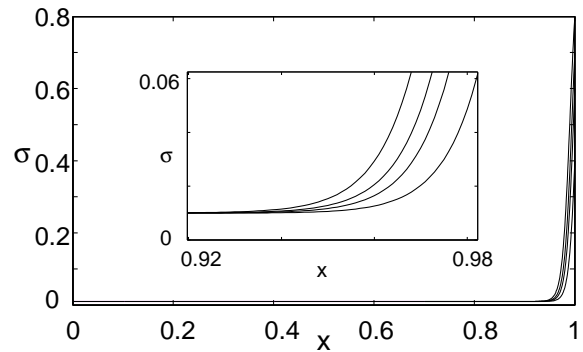}
    \caption{Density profiles, $\sigma$, plotted for $m=0.4$ and  same 
boundary conditions as that of figure \ref{fig:fullden1}.   
 The lower most profile corresponds to $(\rho_r,\sigma_r)=(0.2,0.4)$.
The rest follow the same order as specified in figure \ref{fig:fixedmpt4}.
All the plots are obtained with $\epsilon=0.0062$.}
\label{fig:fullden2}
  \end{center}
\end{figure}

\begin{figure}[htbp]
  \begin{center}
   \includegraphics[width=3.5 in,height=2.5 in, clip]{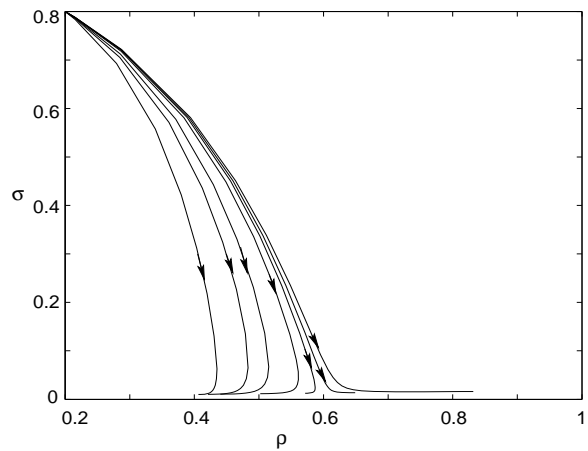}
    \caption{Phase-plane diagram on the $\rho-\sigma$ plane for different 
values of $m$. Trajectories are drawn with initial condition  		 
$\rho(\tilde x=0)=0.2$ and $\sigma(\tilde x=0)=0.8$. The arrows 
indicate the direction of more negative value of $\tilde x$. 
Values of $m$ for various  lines as we move towards the outermost line 
are $0.45$, $0.58$, $0.64$, $0.70$, $0.72$, $0.73$ and $0.74$. 
The flow of the trajectories changes at $m\approx 0.725$.}
\label{fig:naivepp}
  \end{center}
\end{figure}

\begin{figure}[htbp]
  \begin{center}
\includegraphics[width=3.5 in,height=2.5 in,clip]
{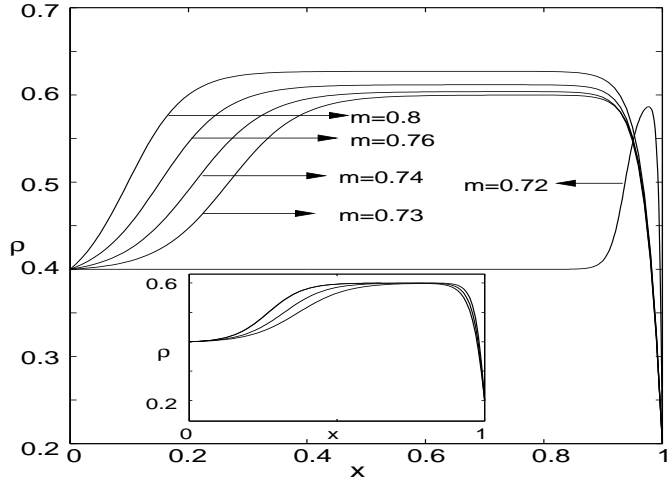}
    \caption{Density profiles for different values of $m$  with 
boundary conditions $(\rho_l=0.4,\sigma_l=0.01)$ and 
$(\rho_r=0.2,\sigma_r=0.8)$. Values of $\epsilon$ for the 
density profiles with $m=0.72$ and $0.73$ are $.0.0025$ and $0.012$, 
respectively. The rest of the profiles are obtained with $\epsilon=0.0125$. 
Inset shows the change  in the shape of the density profile for $m=0.73$ 
 as  $\epsilon$ is  changed. Values of $\epsilon$ as one  moves towards 
the outermost curve are $0.016$, $0.014$ and $0.012$. }
\label{fig:fullden3}
  \end{center}
\end{figure}

The eigenvalues show that of the  two physically meaningful 
fixed points, $(\rho_1^*,\sigma_1^*)$ is an unstable fixed point
 and $(\rho_2^*,\sigma_2^*)$ is a  saddle one.  
Figure \ref{fig:fixedmpt4} shows the flow of 
the phase plane trajectories toward the unstable fixed point 
$(\rho_1^*=0.4, \sigma_1^*=0.01)$ in the $\tilde x\rightarrow -\infty$ limit 
for various initial values.  Although, one can attempt 
to draw the flow trajectories intuitively,  flow  trajectories 
in the figures are obtained by numerically solving
equations (\ref{bl1}) and (\ref{bl2}) with values of 
$c_1$ and $d_1$ determined 
using equations (\ref{const1}) and (\ref{const2}).
Figure \ref{fig:fixedmpt4} indicates that for values of 
$(\rho_r,\sigma_r)$ same as the initial conditions of figure 
\ref{fig:fixedmpt4} and for $(\rho_l=0.4,\ \sigma_l=0.01)$, the density 
profiles must have a  boundary layer part 
near $x=1$ merging to constant 
bulk profiles with $\rho_b=0.4$ and $\sigma_b=0.01$. This bulk part 
satisfies the boundary condition at $x=0$.  
Numerical solutions of the  hydrodynamic equations 
for these boundary conditions 
are shown  in figures \ref{fig:fullden1} and \ref{fig:fullden2}.
Figure \ref{fig:fixedmpt4} shows that the trajectory with initial 
condition $(\rho,\sigma)=(0.2,0.8)$ comes to  close proximity of 
the saddle fixed point. This is reflected in  the boundary layer 
 of the density profile plotted in figure \ref{fig:fullden1} 
 with boundary condition, $(\rho_r,\sigma_r)=(0.2,0.8)$. 
We next study how a trajectory with a given initial condition changes 
with $m$.  
Figure  \ref{fig:naivepp} shows   
phase plane trajectories  with   initial 
condition $(0.2, 0.8)$ for different values of $m$. 
From this figure, it appears that there exists a  special value 
of $m$, say, $m_c$, for  which the trajectory is a separatrix 
which approaches the saddle fixed point. Our numerical solutions show 
$m_c\approx .725$.
Hence, for $m<m_c$  and   $(\rho_r=0.2,\ \sigma_r=0.8)$, 
the bulk-densities are  
$\rho_b=0.4$ and $\sigma_b=0.01$. At $m=m_c$, the bulk density value
 discontinuously 
changes to $\rho_b\approx 0.597,\ \sigma_b=0.0125$. 
Naturally, now the  boundary layers 
are present  at both the boundaries   to satisfy respective  
boundary conditions. These boundary layers are the two separatrices
that approach(emerge) to(from) the saddle fixed point.  
Density profiles with different values of $m$ 
 are presented in figure \ref{fig:fullden3}. The inset in figure 
\ref{fig:fullden3} shows 
that the boundary layer near $x=0$ becomes 
sharper as the value of $\epsilon$ is reduced.
The fact that, for $m>m_c$, the bulk density continues to be 
described by a saddle fixed point is 
clear from the boundary layers at both 
the ends.

\begin{figure}[htbp]
  \begin{center}
\includegraphics[width=3.5 in,height=2.5 in,clip]
{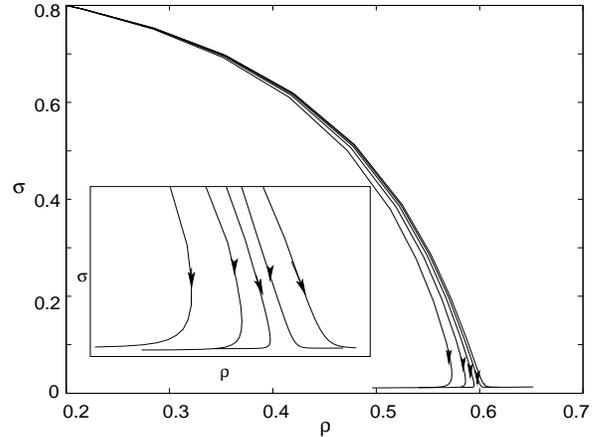}
    \caption{Phase-plane diagram for the boundary layer  
equations with non-linear regularization terms. Values of $m$ 
as one approaches the outermost line are $0.63$, $0.64$, $0.645$, 
$0.648$ and $0.65$. Trajectories 
are obtained with initial  conditions $\rho(\tilde x=0)=0.2$ and 
$\sigma(\tilde x=0)=0.8$.  Arrows on the line 
indicate direction along which $\tilde x$ becomes more negative. 
A zoomed version of the same figure is shown in the inset.}
\label{fig:phaseplane3}
  \end{center}
\end{figure}

Boundary layers for the equations in (\ref{cont1}) 
and (\ref{cont2}) satisfy the 
following first order equations, 
\begin{eqnarray}
(1-m \sigma)\frac{d\rho}{d\tilde x}=
\rho(1-\rho)(1-m \sigma)+c_2\label{nonlin1} \\
(1-m\rho)\frac{d\sigma}{d\tilde x}=
\sigma(1-\sigma)(1-m \rho)+d_2,\label{nonlin2}
\end{eqnarray}  
where, as before, $c_2$ and $d_2$ are two integration 
constants.
These nonlinear  terms do not disturb 
the boundary layer fixed points and since $0\le m\le 1$ and
  $0\le \rho,\sigma\le 1$, the stability properties of the 
fixed points also remain the  same as before. These new factors only 
introduce minor, quantitative 
 changes  in the flow trajectories and in the value 
of $m_c$. The flow  trajectories for these differential 
equations  with the initial condition, $(\rho,\sigma)=(0.2,0.8)$,
and different values of $m$  are shown 
in figure \ref{fig:phaseplane3}. 
As a consequence, 
here also the density profiles 
indicate   a discontinuous change in the bulk density 
as  the value of $m$ is increased.  
This result contradicts earlier results of  
\cite{popkov2} where, from numerical solutions of the 
hydrodynamic equations 
(\ref{cont1}) and (\ref{cont2}), it has been concluded  
that increasing the value of  $m$ does not cause a sudden 
change in the value of the bulk density.  The present  analysis 
brings out the mathematical reason behind 
such a discontinuous change. Our numerical solutions 
of the hydrodynamic equations obtained using MATLAB 
 appear to be consistent with this observation.

\section{Model-2: Two-lane process with particle exchange}

\begin{figure}[htbp]
  \begin{center}
   \includegraphics[width=3.0 in,height=1.0 in, angle=0,
 clip]{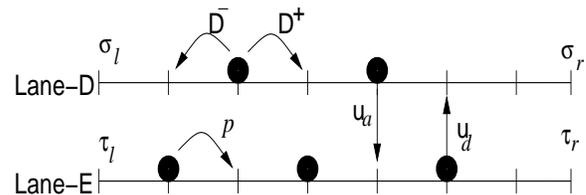}
    \caption{Particle dynamics on the two lanes for model-2.}
\label{fig:figmodel2}
  \end{center}
\end{figure}

Here we consider a system  of two lanes with  hopping of particles 
from one lane to the other \cite{evansprl,evans2}. 
Particle dynamics on the lanes are governed 
by the following rules (See  figure (\ref{fig:figmodel2})).
(a) On one of the lanes, say lane-E, 
particles undergo asymmetric simple exclusion process. This 
implies that particles hop to the neighboring site 
in a specific direction provided it is empty. We assume the 
hopping rate to be $p$ here.

(b) On the other lane, say lane-D, particles are not subjected to 
exclusion interaction and  have 
biased diffusion with hopping rates to left and  right being $D^-$ 
and $D^+$ respectively.  

(c) A particle at the $i$th site on lane-E  can hop to 
the $i$th site of lane-D at  rate $u_d$ and the reverse process 
i.e. an attachment of a particle to lane-E  from lane-D 
 can take place at rate, $u_a$,  provided the target site on 
lane- E is empty. 

In the steady state, the continuum limit of the statistically 
averaged master equations is 
\begin{eqnarray}
{\epsilon}D_\tau\frac{\partial^2 \tau}{\partial x^2} 
-\frac{\partial J_\tau}{\partial x}
-D \tau+A\sigma(1-\tau)=0, 
\label{tau}\\
\epsilon D_\sigma\frac{\partial^2 \sigma}{\partial x^2} -
\frac{\partial J_\sigma}{\partial x}+
D \tau-A\sigma(1-\tau)=0,
\label{sigma}
\end{eqnarray}
where $D_\tau=p/2$, $D_\sigma=(D^++D^-)/2$, $D=u_d N$, $A=u_a N$ and  
$\epsilon=1/N$. Here $\tau(x)$ and $\sigma(x)$ are the average densities 
of particles on the lanes with ASEP and biased diffusion 
respectively  with currents on these lanes being  
\begin{eqnarray}
J_\tau=p \tau(1-\tau), \ \ {\rm and} \ \ J_\sigma=v\sigma.
\end{eqnarray}
Here, $v=D^+-D^-$ denotes the net average velocity of particles 
along lane-D. In addition, we assume that the 
particle reservoirs impose the boundary conditions 
$(\tau(x=0)=\tau_l,\sigma(x=0)=\sigma_l)$ and 
$(\tau(x=1)=\tau_r,\sigma(x=1)=\sigma_r)$.

Constant bulk profiles are expected if the terms due to 
particle exchange 
between the lanes disappear altogether \cite{evansprl,evans2}. 
This happens under the condition 
\begin{eqnarray}
\sigma=\frac{D}{A} \frac{\tau}{1-\tau}.\label{constprof}
\end{eqnarray} 
In the following subsection, we consider this model 
of  constant bulk profile.  Although the phase-diagram 
for the constant profile case has been obtained earlier \cite{evans2}, 
this analysis allows us to see how 
the  maximal and  minimal current phases and upward and downward 
shocks  appear naturally due to 
the flow of the boundary layer solution toward specific fixed points.
This study also provides an ideal  platform to compare this case  
with that  of a non-constant bulk profile  considered  in the 
next subsection.   Therefore, apart from examples with 
different boundary conditions, the next subsection, also 
contains a part in which major differences from the 
constant bulk profile case 
are  discussed.

\subsection {Constant bulk profile}
\begin{figure}[htbp]
  \begin{center}
   \includegraphics[width=3.0 in,height=2.5 in, angle=0,
 clip]{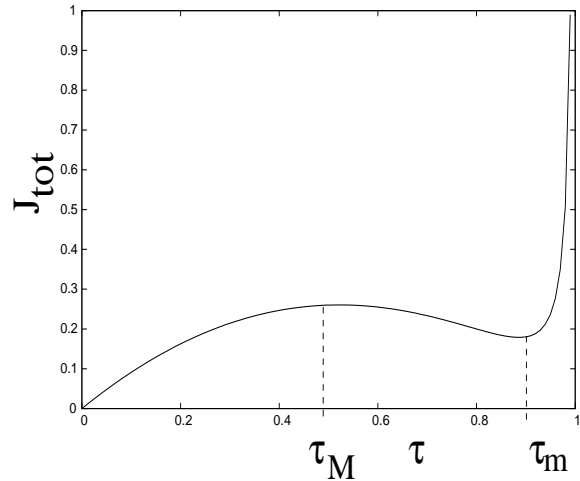}
    \caption{$J_{\rm tot}$ is plotted with $\tau$. For this plot,
$v=p=1$ and $D_{\rm ad}=0.01$.}
\label{fig:jtotal}
  \end{center}
\end{figure}

Adding equations (\ref{tau}) and (\ref{sigma}), and using 
 (\ref{constprof}), we obtain  the following 
 steady state equation  for $\tau$ 
\begin{eqnarray}  
&& \epsilon\frac{\partial}{\partial x}
[D_\tau \frac{\partial \tau}{\partial x}+
D_\sigma D_{\rm ad} \frac{\partial}{\partial x}
\left(\frac{\tau}{1-\tau}\right)]-
\frac{\partial}{\partial x} \nonumber\\
&& [v D_{\rm ad} \frac{\tau}{(1-\tau)}+ p \tau(1-\tau)]=0,\label{bleqn2}
\end{eqnarray} 
which should be solved in the presence of the boundary conditions,
$\tau(x=0)=\tau_l$ and $\tau(x=1)=\tau_r$. Here, $D_{\rm ad}=D/A$.
Terms 
 within the second square bracket of equation
 (\ref{bleqn2}) together give 
the total current $J_{\rm tot}=J_\tau+J_\sigma$ on the two lanes. 
Special values of $\tau$, $\tau_M$ and $\tau_m$, corresponding to 
maximum and minimum of $J_{\rm tot}$ (see figure \ref{fig:jtotal}) 
play an important role in 
the phase-plane analysis. 

\begin{figure}[htbp]
  \begin{center}
   \includegraphics[width=3.0 in,height=3.5 in, angle=270,
 clip]{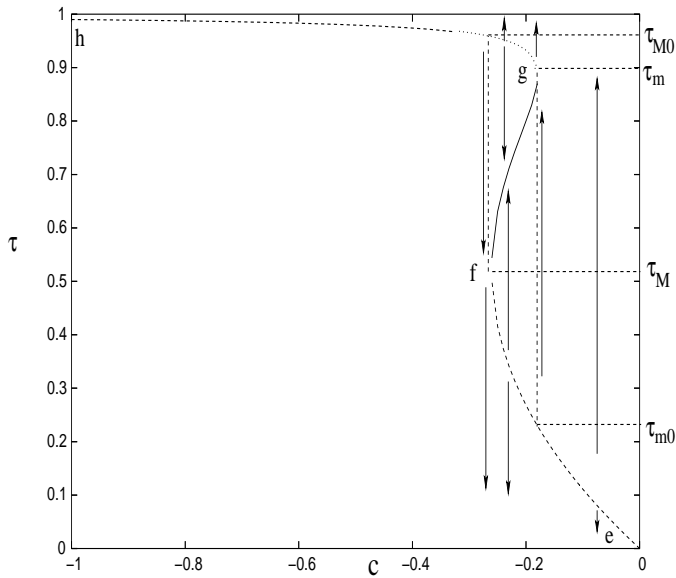}
    \caption{Fixed point diagram on the $c-\tau$ plane for 
the parameter values $v=1$, $p=1$ and $D_{\rm ad}=0.01$. 
The arrowed vertical  lines  show the stability properties 
of various fixed point branches. Three fixed point branches, 
ef, fg and gh are referred in the text
 as lower, middle and upper branch respectively.  }
\label{fig:pplane_diffus}
  \end{center}
\end{figure}

Once again, the  boundary layers 
are the solutions of the second order differential equation
  (\ref{bleqn2}) which after one integration appears as
\begin{eqnarray}
&& \frac{\partial \tau}{\partial \tilde x}[D_\tau+
D_\sigma D_{\rm ad} \frac{1}{(1-\tau)^2}]-
[v D_{\rm ad} \frac{\tau}{1-\tau}+\nonumber\\
&& p \tau(1-\tau)]=c,\label{bleqn3}
\end{eqnarray} 
Here $c$ is the  integration constant and $\tilde x=\frac{x-x_0}{\epsilon}$. 
Saturation of the boundary layer to the bulk density, $\tau_b$, requires
\begin{eqnarray}
c=-v D_{\rm ad} \frac{\tau_b}{1-\tau_b}-p \tau_b(1-\tau_b).
\end{eqnarray}
Since $0\le\tau_b\le 1$, the value of $c$ lies
 in the range $-\infty<c<0$. For a single differential equation as 
(\ref{bleqn3}), one can have a two-dimensional fixed point diagram 
as shown in figure \ref{fig:pplane_diffus}. This is an additional 
advantage of the present system over the previous one.  

 The basic  principle for finding the density profiles using the 
fixed point diagram is that  
the shocks or the boundary layers in the profiles 
correspond to vertical straight lines on the fixed point diagram
  moving from 
one fixed point branch to the other following the 
stability  properties. For example, an upward shock in the 
bulk of the density profile may connect a low-density at 
the lower fixed point 
branch to a higher density in the middle fixed point branch. 

(a)  Given this rule, the density profile can have   localized  shocks 
under the following conditions. 
(i) $\tau_{m0}<\tau_l<\tau_M$, $\tau_M<\tau_r<\tau_m$ 
 and  $c(\tau_l)=c(\tau_r)$.

\begin{figure}[htbp]
  \begin{center}
   \includegraphics[width=3.0 in,height=3.5 in, angle=270,
 clip]{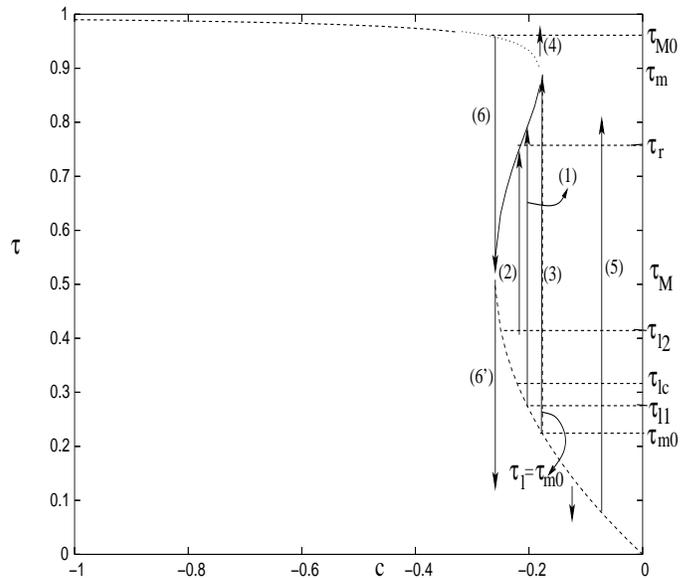}
    \caption{Fixed point diagram of figure \ref{fig:pplane_diffus}. 
Vertical lines are numbered in order to refer these in the text in 
the context of different boundary layers and shocks.}
\label{fig:pplane-diffus1}
  \end{center}
\end{figure}

Pictorially, an upward  shock is possible if 
 $\tau_l$ located in the lower fixed point 
branch can be connected to $\tau_r$ in the middle branch through 
only a vertical straight line. This constraint is  stated mathematically 
through the above equality satisfied by $c$. 
Let us assume that for a given $\tau_r$ in 
the middle branch, the localized shock appears if $\tau_l=\tau_{\rm lc}$ 
(See figure \ref{fig:pplane-diffus1}.). 
Now for $\tau_l=\tau_{l1}<\tau_{\rm lc}$,  
the flow behavior  allows  a density profile 
with a  bulk part   satisfying the boundary condition at $x=0$ and 
a boundary layer near $x=1$. This kind of boundary layer is  
represented by vertical line (1) in figure \ref{fig:pplane-diffus1}. 
In this case, it is not 
possible to have a  boundary layer at 
$x=0$  since that would imply a vertical line passing 
through $\tau_l$ and  going further upward after crossing $\tau_r$. 
Although such a line  passes  through both $\tau_l$ and $\tau_r$, 
it cannot represent a bulk profile. 
Similarly, for $\tau_l=\tau_{l2}>\tau_{\rm lc}$, a profile,
 with bulk satisfying 
the boundary condition at $x=1$ and a 
boundary layer near $x=0$ represented by line (2) in figure
\ref{fig:pplane-diffus1}, is the only solution. 


(ii) A  downward   shock in the bulk profile can be present  
when $\tau_m<\tau_l<\tau_{M0}$ and $\tau_M<\tau_r<\tau_m$ and 
$c(\tau_l)=c(\tau_r)$. 
In this case, a bulk profile of value $\tau_l$ at left is joined to a
 bulk density of  value $\tau_r$ through  a downward  shock. 
 

(iii) Another very special line on  which localized 
upward shocks can be  present is given by $\tau_r>\tau_m$ and 
$\tau_l=\tau_{m0}$.  The shock here joins $\tau_l$ to the minimum current 
density $\tau_m$. The density profile at  $x=0$ starts with a constant 
value $\tau_l$ and then reaches another constant part $\tau_m$ through a 
shock, represented by the vertical line (3) in 
figure \ref{fig:pplane-diffus1}, 
and finally satisfies the boundary condition $\tau_r$ through a boundary 
layer, represented by the vertical line (4) in the figure. 

Approach of a shock or a boundary layer to the bulk density is governed 
by equation (\ref{bleqn3}). The 
variation of a  small perturbation $\delta\tau$  near 
 the saturation to a  bulk density $\tau_b$ is given by 
\begin{eqnarray}
\frac{d\delta\tau}{d x} f(\tau_b)=[\frac{v D_{\rm ad}}{(1-\tau_b)^2}+
p(1-2\tau_b)]\delta\tau,
\end{eqnarray} 
where $f(\tau)=D_\tau+D_\sigma D_{\rm ad} \frac{1}{(1-\tau)^2}$.
Thus the  approach of a boundary layer to the bulk  is in general 
exponential except at  special bulk values 
$\tau_b=\tau_m,\ \tau_M$. The 
length scale associated with the exponential 
approach to the bulk diverges at these special 
values \cite{evans2,hager}.  

(b) Boundary conditions for  which a minimal current phase may appear:
The flow lines clearly show that for 
 $\tau_r>\tau_m$  and $\tau_{m0}<\tau_l<\tau_m$, a  
shock connecting  $\tau_l$ to the middle branch is not helpful for meeting 
the boundary condition at $x=1$. 
In this case the only option is to have a 
boundary layer at $x=0$ connecting $\tau_l$ to $\tau_m$ and then a constant 
 bulk profile of  density $\tau_m$ followed by a boundary layer at $x=1$. 
Such a boundary layer must be represented by line (3) in figure 
\ref{fig:pplane-diffus1}. 
Note that this is the only possibility since any other parallel 
vertical line with $c>c(\tau_m)$ will correspond to  an unphysical 
solution.
This explains how a minimal current density profile becomes 
an obvious solution for boundary densities in this region.
For the opposite case, $\tau_l<\tau_{m0}$ and same $\tau_r$, 
the density should have a constant value  $\tau_l$ with a boundary layer 
at $x=1$ represented by line (5) in figure \ref{fig:pplane-diffus1}.   

(c) Boundary conditions for  which a maximal current phase may appear:
For $\tau_M<\tau_l<\tau_{M0}$ and $\tau_r<\tau_M$, the only 
possibility is to 
have a boundary layer at $x=0$ represented by line (6) in figure 
\ref{fig:pplane-diffus1}. This line meets the fixed point curve at 
$\tau_M$ which continues as the bulk density. 
The boundary condition at $x=1$ is 
met by another downward vertical line (6'). It is clear that 
if $\tau_l=\tau_{M0}$, the density profile can have a localized 
downward shock
of height $\tau_{M0}-\tau_M$. The bulk density will no longer 
have a value $\tau_M$ if $\tau_l>\tau_{M0}$. In that case, 
the only possible shape for the density profile is a flat 
profile of density $\tau_l$ followed by a downward 
boundary layer  satisfying the boundary condition $\tau_r$.

\subsection{Non-constant bulk profile}

In this  section, we illustrate how a similar analysis as that of the 
previous sub-section can be extended to a case of a non-constant bulk 
profile. In order to produce  a non-constant bulk profile, we 
artificially add a term to equation (\ref{bleqn2}). This kind of 
term has appeared earlier  in the steady-state equation ASEP with 
Langmuir kinetics \cite{parameg}. However, for this model, this term 
only has a 
 mathematical implication of producing a non-constant bulk profile. 
The final steady state equation for the density $\tau$ is
\begin{eqnarray}  
&& \epsilon\frac{\partial}{\partial x}[D_\tau 
\frac{\partial \tau}{\partial x}+
D_\sigma D_{\rm ad} 
\frac{\partial}{\partial x}\left(\frac{\tau}{1-\tau}\right)]-
\frac{\partial}{\partial x}[v D_{\rm ad} \frac{\tau}{(1-\tau)}+\nonumber\\
&& p \tau(1-\tau)]+\Omega (1-2 \tau)=0.\label{bleqnnonconst1}
\end{eqnarray}   
 Ignoring the 
second order derivative term in (\ref{bleqn2}), one may consider
 a simplified equation, valid at $O(\epsilon^0)$, as
\begin{eqnarray}
[\frac{vD_{\rm ad}}{(1-\tau)^2}+
p(1-2\tau)]\frac{\partial \tau}{\partial x}+\Omega(2\tau-1)=0.\label{outer}
\end{eqnarray} 
As mentioned earlier, the solution of this equation describes  
the bulk profile.  
In principle, one can solve this equation explicitly to 
find the bulk parts of the density profile. However, as we shall show, 
the shape of the profile can be predicted without explicitly solving 
this or the boundary layer equation. 

 After a brief comparison between the two cases 
with constant and non-constant bulk profiles, 
we discuss  a few examples with  different boundary 
conditions.  To show that our approach gives the right 
profile, we illustrate these examples 
with typical density profiles obtained by  
solving the hydrodynamic equation numerically. 
We hope that these ideas can be 
implemented more generally 
for all other boundary conditions that 
are not discussed here.

{\bf Comparison between the cases with 
constant and non-constant bulk profiles:}

\begin{figure}[htbp]
  \begin{center}
   \includegraphics[width=3.0 in, height=3.5 in, angle=270,
 clip]{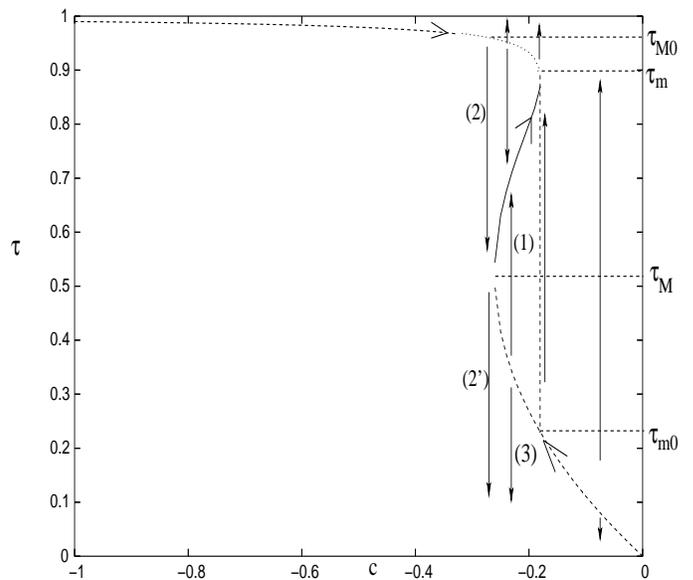}
    \caption{Fixed point diagram of figure \ref{fig:pplane_diffus}.
Vertical lines are numbered in order to refer these in the text in the 
context of different shocks and boundary layers.}
\label{fig:pplane-diffus2}
  \end{center}
\end{figure}

The major difference from the previous analysis is that 
for a flat bulk profile, there is no variation in the density, once 
the boundary layer merges to a  fixed point. The density in the bulk 
remains constant at this fixed point value. This need not be the case here
and, in general, the density varies along various fixed point branches
after the boundary layer merges to a fixed point. Thus  
the entire density 
profile can be predicted, by knowing only the slope of the density 
profile in the bulk and following the vertical arrowed lines for the 
boundary layer parts.  The slope of the bulk profile can be 
determined from (\ref{outer}) and the arrows on the fixed point 
branches in figure \ref{fig:pplane-diffus2} represent increasing or 
decreasing nature  of the bulk density with $x$.
Since for a flat profile,  the density at the bulk 
has to remain constant at $\tau_l$, $\tau_r$ or at 
very special values 
like, say, $\tau_M$, the possibility of seeing localized shock here 
 is restricted. 
The shock appears only when these  values are such that 
they  can be connected 
through vertical lines in the interior of a lane. 
Such limitations are not present here since the varying bulk density 
may reach certain values 
which can be connected by vertical lines in the bulk. 
It is because of this that 
 seeing localized shocks over a region in the phase 
diagram becomes more likely, in general, whenever the bulk density 
is not constant and its  slope is appropriate for supporting a shock.
Such a phase diagram with a wide region of localized shock 
was reported initially in \cite{parameg}. 
    
\begin{figure}[htbp]
  \begin{center}
   \includegraphics[width=3.5 in, height=3.0 in, angle=0,
 clip]{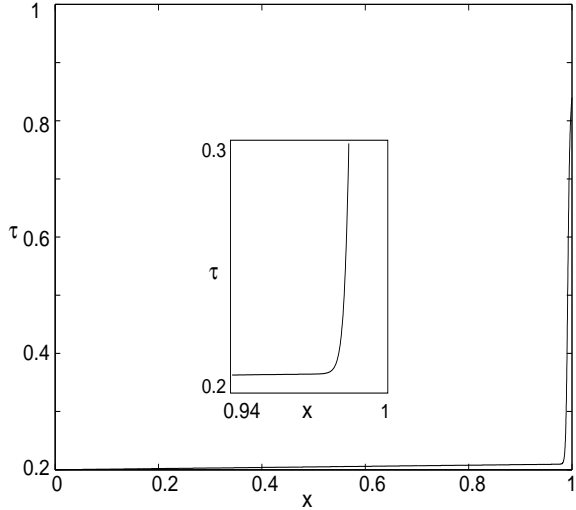}
    \caption{Density profile for $\tau_l=0.2$ and $\tau_r=0.84$. 
Values of other parameters are $\epsilon=0.002$, $v=p=1$, 
$D_{\rm ad}=0.01$, $\Omega=0.01$  and $D_\sigma=1$.}
\label{fig:densprofnc1}
  \end{center}
\end{figure}

 \begin{center}
 $\mathbf{\tau_l<\tau_{m0}}$ {\bf and}  ${\mathbf {\tau_M<\tau_r<\tau_m:}}$ 
\end{center}
There cannot be a 
boundary layer at $x=0$ since that would correspond to a  vertical flow 
line passing through $\tau_l$ and the density profile will not 
correspond to a physically meaningful solution. However, a bulk profile 
satisfying the boundary condition at $x=0$ is possible. This is an 
increasing profile along the lower fixed point branch. From this 
lower branch, there are 
two ways to satisfy the  boundary condition at $x=1$: 
(a) through a shock (represented by a vertical line in the phase plane) 
connecting the low-density bulk part in the lower branch
 to a high-density bulk part in the middle branch. 
This latter bulk part,  with a positive slope (see the upward arrow 
along the middle fixed point branch)  finally
satisfies the boundary condition at $x=1$. 
(b) through a boundary layer  (represented by a vertical line 
in the phase plane) that connects the bulk density  in the lower branch 
to a density in the middle branch. The  
boundary layer satisfies the boundary 
condition at $x=1$ before it saturates to  the other fixed point. 
Figure \ref{fig:densprofnc1} shows a density profile following option (b).

\begin{center}
$\mathbf{\tau_{m0}<\tau_l<\tau_M}$ {\bf and} 
$\mathbf{\tau_M<\tau_r<\tau_m}$:
\end{center}

\begin{figure}[htbp]
  \begin{center}
   \includegraphics[width=3.5 in, height=3.0 in, angle=0,
 clip]{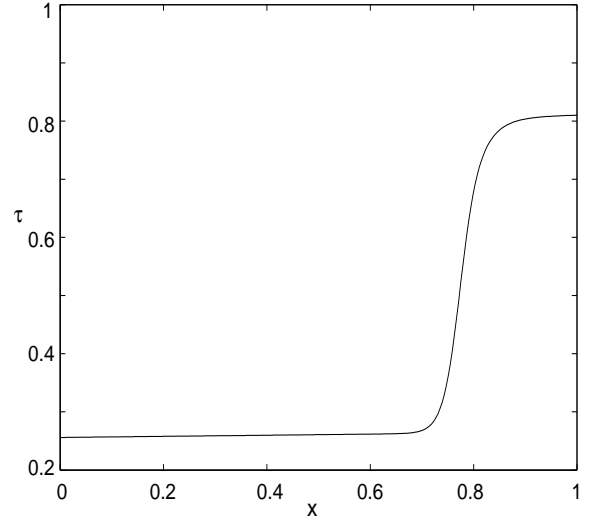}
    \caption{Density profile with a shock for 
 $\tau_l=0.256$,  $\tau_r=0.81$ and  
$\epsilon=0.015$. Values of other parameters are same as those of figure 
\ref{fig:densprofnc1}}
\label{fig:densprofnc2}
  \end{center}
\end{figure}

An increasing  bulk density satisfying the boundary condition at $x=0$
may be connected to another bulk part in the middle branch through a 
shock represented by the vertical line (1) in 
figure \ref{fig:pplane-diffus2}. This is similar to option (a) of the previous 
example. Such a density profile appears as shown in figure 
\ref{fig:densprofnc2}.
There are still two other options  each with an  upward boundary layer.
One possibility is that an upward boundary layer 
at $x=0$ meets the middle 
fixed point branch and  a bulk part continuing 
 from there  finally satisfies 
the boundary condition at $x=1$. The other possibility is that a 
bulk part with positive slope 
satisfies the boundary condition at $x=0$  and  an upward boundary 
layer following the bulk satisfies the boundary condition at $x=1$.  

\begin{center}
 ${\mathbf {\tau_r<\tau_{M}}}$
 and ${\mathbf{\tau_M<\tau_l<\tau_m:}}$
\end{center}
This is a very unique situation for the following reasons. 
Figure  \ref{fig:pplane-diffus2} 
shows that there is only one route on the fixed point  diagram 
that the density profile must follow to satisfy both the boundary 
conditions.  At $x=0$, there must be a 
boundary layer of negative slope 
(represented by a vertical line  (2)) merging to  density 
$\tau_M$. This should be followed by another downward profile 
(represented by (2')) that satisfies the boundary condition 
 at $x=1$.
The entire 
density profile is thus described by only two 
vertical lines (2) and 
(2'). A 'bulk' like part of  almost constant 
value ($\approx \tau_M$) does appear in the density profile (see figure 
{\ref{fig:densprofnc3}). 
However, this is not a real bulk profile since 
$\tau_M$ is not a solution of (\ref{outer}). It  looks like 
a bulk density since the slopes of the vertical lines (2) and (2') 
vanish at $\tau=\tau_M$ and hence the profile  looks 
almost  flat.  $\tau=\tau_M$ is in fact  a point of inflection of 
the density profile. In this sense, there is a subtle difference 
between the maximal current phase seen here and in the simple ASEP model  
for which the bulk-density value, $1/2$, 
is  a solution of the 
corresponding outer equation  \cite{dualjphys}.

\begin{figure}[htbp]
  \begin{center}
   \includegraphics[width=3.5 in, height=3.0 in, angle=0,
 clip]{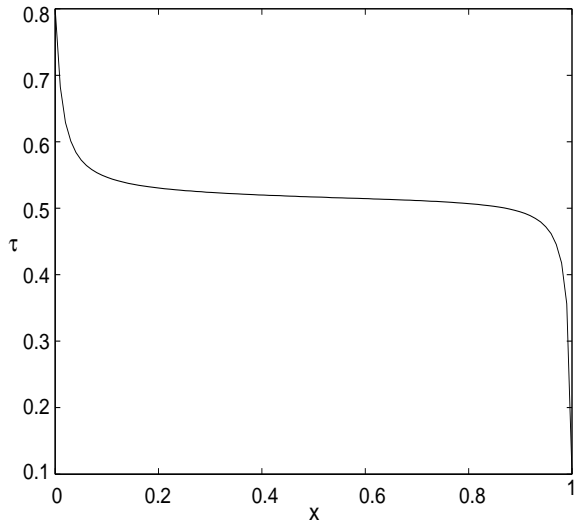}
    \caption{Density profile for $\tau_l=0.8$, $\tau_r=0.1$
and $\epsilon=0.005$. Values of other parameters are same as those of
figure \ref{fig:densprofnc1}.}
\label{fig:densprofnc3}
  \end{center}
\end{figure}

\begin{center}
 ${\mathbf {\tau_l,\tau_r < \tau_{m0}}}$ {\bf and} 
${\mathbf {\tau_l>\tau_r:}}$
\end{center}
The bulk profile must satisfy the boundary condition at $x=0$ 
and increase along the lower branch 
 and then satisfy the other boundary condition 
through a boundary layer of negative slope at $x=1$, 
represented by a vertical line (3) in the figure 
\ref{fig:pplane-diffus2}. A typical density 
profile appears as shown in figure \ref{fig:densprofnc4}.

\begin{figure}[htbp]
  \begin{center}
   \includegraphics[width=3.5 in, height=3.0 in, angle=0,
 clip]{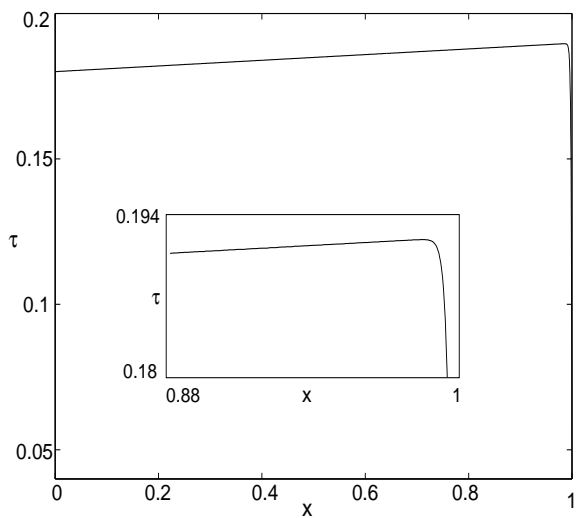}
    \caption{Density profile for $\tau_l=0.18$,  
$\tau_r=0.04$ and $\epsilon=0.002$. Values of  other parameters 
are same as those of figure \ref{fig:densprofnc1}.}
\label{fig:densprofnc4}
  \end{center}
\end{figure}

\section{conclusion}
We have shown how the phase-plane behavior of the boundary-layer equation 
can be used to understand the shape of the steady-state 
density profile of multi-lane driven exclusion processes. We have 
considered two distinct multi-lane  processes. In the first model, the lanes 
do not exchange particles but the particle dynamics on the two lanes 
are coupled  since 
hopping of particles on one lane is affected by the particle occupancies 
on the neighboring lane. In the hydrodynamic approach, the 
 boundary layers of the  density profiles on the two lanes 
are described by coupled nonlinear equations.  Using phase plane 
trajectories, 
we show that, as the interaction between the lanes is increased, 
the bulk density, for certain boundary conditions,  may 
increase discontinuously  due to the influence 
of a saddle fixed point. This has been done using linear  
phenomenological regularization 
terms for the differential equations. Since this method does not 
require explicit solution of  the density profile, it can be easily 
extended to other system  of  equations with 
nonlinear  regularizing terms. One such case with nonlinear 
regularizing terms has been discussed here.   
The second model involves two  lanes
 which can mutually exchange 
particles. In addition,  particles in one lane go through
a driven exclusion process and those  on the other lane have 
biased diffusion without any exclusion constraint. We consider 
two variants of this model; one with a constant bulk density profile
and another with a non-constant bulk profile. The fixed point diagram shows
saddle node bifurcations of the fixed points 
of the boundary-layer equation. 
For constant bulk profile, the fixed point diagram  allows us to 
predict 
the density profiles and the location of the boundary layers 
uniquely  under various boundary conditions. 
For non-constant bulk-profile, one requires an additional information 
regarding the slope of the bulk solution. Some of the useful features
of the method  are as follows.  With the required information about the 
bulk profile, one can exactly 
predict under which condition   boundary layer or shock will have 
positive or negative slope. Similar analysis can be done  with ease 
for different  parameter values which may reverse  the slopes of the 
bulk solutions on some of the fixed point branches 
or change the fixed-point diagram significantly.
Analysing the two cases of constant and non-constant profiles, 
it becomes transparent as why 
localized shocks appear only under restricted 
boundary conditions for a flat profile but over a wide range of  
boundary conditions for a non-constant profile.

We have analysed the first 
 model for only some specific  
 boundary conditions and we have not made any attempt 
to explore the phase diagram although the method
can be used in a similar manner 
for other boundary conditions also. There are 
several aspects which may make the 
analysis interesting. One, for example, is  the 
presence of  more number of 
physically acceptable fixed points. This, apart 
from leading to interesting phase portraits, 
 may  reveal the role of various eigenvalues 
in deciding the shape of the density profile.
Second, is the  presence of imaginary 
eigenvalues which may be responsible for 
oscillatory density profiles. Whether the multi-lane 
problem considered here has this kind of complexity 
can be explored through a detailed analysis 
for the full phase-diagram. This work is in progress.

For the second model, the condition relating the 
two densities on the two lanes simplifies the problem 
in two ways. First, it leads to constant density profile and second, 
due to this relation, it becomes sufficient to consider 
hydrodynamic equation for a single density variable. 
In the absence of  such a condition,  individual 
boundary layer equations look apparently decoupled at the lowest 
order. The boundary layer analysis for the two densities 
 would, however, be coupled due to the coupled nature 
of the bulk equations.  This may give rise to an interesting 
boundary layer analysis not encountered earlier.


\end{document}